\documentclass[vecphys]{svmult}

\usepackage{makeidx}         % allows index generation
\usepackage{graphicx}        % standard LaTeX graphics tool
\usepackage{multicol}        % used for the two-column index
\usepackage[bottom]{footmisc}% places footnotes at page bottom
\makeindex             % used for the subject index
\begin{document}

\title*{The double life of the $X$ meson} 
%and its {\it similes}}
\author{A.D. Polosa}
\institute{INFN Sez. di Roma 1, P.le A. Moro 2, I-00185 Roma, Italy.
\texttt{antonio.polosa@roma1.infn.it}
}
\maketitle

Three years have passed since the BELLE discovery of the $X(3872)$, and there are still (at least) two competing interpretations of this particle, which resembles a charmonium but behaves in a dramatic different way from it. Is $X$ a molecule of two $D$ mesons or a compact four-quark state?  
Are these two pictures really resolvable?

The quantum mechanical intuition can also lead to more refined pictures: the $X$ could be a sort of breathing mode of a charmonium oscillating into a molecule and back. 
%This not to mention more `fantastic' pictures suggested in a quite vast literature on %the subject.

Other particles have been discovered since then: the 
 $X(3940)$, $Y(3940)$, $Z(3930)$ (amazingly the first two have the same mass and both decay to charmonium but with a different decay pattern to open charm) and
 $Y(4260)$. The latter also decays into $J/\psi$ and could be an `hybrid' particle 
 (two quarks and a constituent gluon), likely the most experiment-proof interpretation so far.

In this talk I will not try to describe all the experimental facts and theoretical ideas, thoroughly reported and commented elsewhere in the literature.
I will rather comment on the first question here raised, namely, how far are molecules, four-quarks or charmonium-molecule oscillating states distinguishable in principle and in the experimental practice? The suspect that the competing scenarios fade into one another could dangerously leave this field in a confused and controversial situation similar to that existing for sub-GeV scalar mesons (and for their super-GeV partners).  
\section{Sewing Quarks}
\label{sec:1}
% Always give a unique label
% and use \ref{<label>} for cross-references
% and 
%{<label>} for bibliographic references
% use \sectionmark{}
% to alter or adjust the section heading in the running head
The prominent decay mode of $X(3872)$ is $X\to J/\psi \rho$. Several studies
conclude that the $X$ cannot be an ordinary $c\bar c$ state, even though the 
$J/\psi$ invokes a charmonium assignation.
The next-to-easy interpretations can be: $(1)$ $X$ is a
 $D\bar D^*$ bound object, with the correct $1^{++}$ quantum numbers. Such a molecule could decay at the same rate to $J/\psi \rho$ and $J/\psi \omega$,
 which is actually what surprisingly happens in nature (this was not a prediction though). This molecule lives for a while, until the two heavy quarks get
 close enough to form a $J/\psi$, leaving the light quarks free to hadronize 
 to a  $\rho^0$. $(2)$ $X$ is a four-quark $c\bar c q \bar q$ meson. The four quarks could be diving in some potential bag but, if so, we should get ${\bf 3}\otimes {\bf \bar 3}\otimes {\bf 3}\otimes {\bf \bar 3}$, i.e., 81 particles.
 This is the obvious problem of exotic structures: a copious number of states 
 is predicted. Moreover such multiquark structures could fall apart and decay at an immense rate (resulting in very broad and experimentally elusive states): at the lowest $1/N_c$ order a propagating state of four quarks in a color--neutral, gauge-invariant combination $q^i\bar q_i q^j\bar q_j$, is indistinguishable from two freely propagating $q\bar q$ mesons. On the other hand, it turns out that quarks (and antiquarks) can be bound in $qq$  ($\bar q \bar q$) diquarks (anti-). As for the color, a diquark is equivalent to an anti-quark and the anti-diquark is equivalent to a quark. A diquark-antidiquark meson is therefore pretty much the same as an ordinary meson, as for the strong force.

A ${\bf \bar 3_c}$ spin zero diquark is antisymmetric in flavor, ${\bf \bar 3_f}$,
because of the Fermi statistics, as long as $q=u,d,s$. Therefore a four-quark system made up of two diquarks  involves ${\bf 3_f}\otimes {\bf \bar 3_f}$ states, 9 states versus the 81 given before (a crypto-exotic) which is much better, 
although $X$ is a $1^{++}$ state and two spin zero diquark cannot do the job.

The $X$ should however contain two $c$ quarks. The heavy quark $Q$ is not indistinguishable from $q$ and spin-spin interactions between an heavy quark and a light one are $O(1/M_c)$, so that, even if non-perturbative dynamics tends to favor the formation of a spin zero diquark (as it has been proved by lattice studies focused on light-light diquarks), an heavy-light diquark can be equally well spin zero or one and its flavor group structure is determined by the light quark only. Again 9 states, but with spin one. The other quantum numbers follow easily.

On the other hand, the number of ways of sewing quarks into a four-quark structure is not exhausted by the  possibilities just described. Two ${\bf 3_c}$ quarks can either attract or repel each other in the 
${\bf \bar 3_c}$ or ${\bf 6_c}$ color channels according to the one-gluon exchange model (which qualitatively reproduces the lattice indications). According to the same model, a ${\bf 6_c}$ diquark and a ${\bf \bar 6_c}$ anti-diquark could form a color neutral object 
%(${\bf 6}\otimes {\bf \bar 6}$, besides ${\bf 8}$ and ${\bf 27}$, contains the singlet ${\bf 1}$) 
with the {\it same} binding energy of the 
${\bf 3_c}-{\bf \bar 3_c}$ diquark--anti-diquark. This object looks like the non-abelian analog of a system of two electrons and two protons in some closed region: an $H_2$ molecule is formed as a result of the binding of two individual hydrogen atoms.

The one-gluon exchange model of the strong interactions in a hadron is just a qualitative oversimplification, yet  it gives the feeling of how the molecule and 
four-quark languages could  dangerously be interchangeable\footnote{
The oscillating $c\bar c$-molecule picture is a smart refinement of the basic molecule description with a stronger adaptability to the sometimes adverse climatic conditions of the experimental situation}.

\section{Tracing differences}
\label{sec:2}
Building four-quark mesons made up of two diquarks  requires 9 states: charged 
$X$'s should be visible, as well as strange $X_s$ states, according to $SU(3)$.
An entire spectrum of these states has been calculated.
Only one neutral non-strange $X$ has been observed so far; similarly no charged partners of the higher mass $X,Y,Z$ are observed. This is usually addressed as the weakest point in the tetraquark picture.
But, (1) even if an attempt to calculate the $X$ mass spectrum in the four-quark picture has been made, it is not at all easy to predict the widths of these nonet states, most of which could turn out to be very broad.
(2) $D\bar D^*$  molecules could as well occur in 9 states, though it seems that binding potentials can be tuned to account for the `reduced' observed spectrum.

At any rate, molecules are very loose bound states: consider for example that $m_{D}+m_{\bar D^*}=3871.3\pm 0.7$~MeV. Then we can expect that the typical size of such a molecule is $r\sim 1/\sqrt{2 M_X E_{\rm bind.}}\sim 3-4$~fm. Charm quarks have to recombine into a $J/\psi$ (kind of $\sim 0.25$~fm object)  starting from a configuration in which they are up to $~ 4$ times the typical range of strong interactions apart.
In the tetraquark picture, instead, the $c$ quarks are as close to each other as two quarks in a standard meson.

A $D\bar D^*$ molecule should have a decay width $X\to D^0 \bar D^0\pi^0$ comparable to the $\Gamma (D^*\to D\pi)\sim 70$~KeV width. This decay mode has been very recently observed to occur at a rate about nine times larger than the $J/\psi\rho$ mode, in bold contradiction with the basic molecular picture where $J/\psi\rho$ was predicted to be by far the dominant one.
The tetraquark $X$ is allowed to decay $X\to D^0 \bar D^0\pi^0$ with a rate almost two times larger than the $J/\psi\rho$.
This experimental fact, if confirmed, seriously challenges both models.

All these semi-quantitative considerations are not definitive in deciding neatly between the two options: molecule or tetraquark?. In many respects one could so far object that the two scenarios seem quite contiguous.
But, in the tetraquark picture the $X(3872)$ has a `double life', two different $X$'s are required (call them $X_l$ and $X_h$) to account for the 
observed isospin violation:
${\cal B}(X\to J/\psi \rho)/{\cal B}(X\to J/\psi \omega)=1.0$.
In what follows I will sketch the latter point.

Consider the states $X_u=[cu][\bar c\bar u]$ and $X_d=[cd][\bar c\bar d]$, where the square parentheses indicate a diquark binding.  The $B^+$ could decay
as $B^+\to K^+X_u$ and $B^+\to K^+X_d$ with some undetermined relative frequency. Let us call $A_1$ and $A_2$ the two decay amplitudes. Data on the production of $X(3872)$ in $B^+\to K^+ X$ reasonably show that only one single state is produced in this channel. Therefore either $A_1>>A_2$  or $A_1<<A_2$. Whatever the actual situation is, a naive quark analysis shows that in  $B^0\to K X$, $A_1$ and $A_2$ would be interchanged: if $X_u$ is 
produced in
$B^+$ decay, then $X_d$ is produced in $B^0$ decay and vice-versa.

Actually, the real $X$'s can be superpositions of $X_u$ and $X_d$. In a standard mixing scheme we can introduce two orthogonal superpositions, $X_l$ and $X_h$,  mixed by an angle $\theta$. The annihilation diagrams describing $u\bar u - d\bar d$ transitions are reasonably quite small at the $m_c$ scale so that we 
expect $\theta$ to be small. 
$X_l$ and $X_h$ are therefore unbalanced superpositions of 
$I=1/2$ and $I=-1/2$ states (at $\theta=\pi/4$, $X_l$ and $X_h$ are $I=0$ and $I=1$ respectively; $X_l$ could, e.g.,  only decay to $\omega J/\psi$)
opening the way to isospin violations in the decays of $X_l$ and $X_h$. On the other hand the $D\bar D^*$ molecule is per-se a single isospin impure state.

$X_l$ and $X_h$ are expected to have a difference in mass, $\Delta M$,
proportional to $(m_d-m_u)$ and inversely proportional to $\cos \theta$ (which can be fixed by decay data). 
Such a mass difference is under experimental study (the mass of the $X$ produced 
in $B^+$ is confronted to the mass of $X$ produced in $B^0$) but the error on data still does not allow to draw a definitive conclusion. A $\Delta M\sim O(1)$~MeV would clearly unveil the double life of $X$ excluding the molecule (and all the way around).

Resolving this molecule-tetraquark dichotomy is not only a matter of taxonomy. Diquarks have an interesting role in QCD. An entire region of the QCD phase diagram in the $(\mu,T)$ plane has been found to exist in a phase of color superconductor where the analogous of the ordinary Cooper pairs are diquarks.
Diquarks also enter in diverse QCD considerations. Just to mention  one, recall for example the argument to explain the limit   $F_2^n/F_2^p\longrightarrow 1/4$ as $x\to 1$ of the DIS structure functions of neutron and proton. Diquarks could also help to explain the fact that the Regge trajectories of mesons and baryons happen to have the same slope.

\section{Counting Quarks}
Obtaining direct experimental evidence that the $X$ is a multiquark object  would certainly be rather useful.
A new method to investigate the quark nature of the $X$ and of all those states missing a clear quark-identikit, like $f_0(500)$, $a_0(980)$, $f_0(980)$..., could be obtained by the analysis of certain heavy-ion collision observables.

A stunning fact emerged at RHIC is that the number of protons divided by the number of pions counted in a $p_\perp$  region
$1.5~{\rm GeV}\leq p_\perp\leq 4~{\rm GeV}$ is $\geq 1$, against any expectation based on fragmentation functions which would predict an opposite pattern.   In such experimental situation, fragmentation is insufficient at producing high $p_\perp$ hadrons. In the standard fragmentation picture, an off-shell parton loses energy via a perturbative `shower' of gluon emissions until the energy scale of 
$\Lambda_{\rm QCD}$ is approached, where the non-perturbative domain opens.
At this stage all the partons formed will get together in separated clusters tending to neutralize the color and generating, via some non-perturbative mechanism, a collection of real hadrons. The energy of the initial parton is repartitioned among all the final state hadrons produced. High $p_\perp$ hadrons in the final state must be originated by very high $p_\perp$ initial partons which, in usual conditions, are not abundantly produced: pQCD spectra are steeply falling with $p_\perp$. Moreover, the standard fragmentation function approach predicts that, for a generic parton $a$, $D_{a\to p}/D_{a\to \pi}\leq 0.2$ in the above $p_\perp$ range.

But, suppose that a rapidly expanding state, overflowed in phase space with partons, is created in a heavy-ion collision. Neighboring partons in phase space could be able to recombine into hadrons whose momenta are just the algebraic sums of the parton momenta involved.
In this case we could state that $[p~{\rm spectrum}]\sim \exp [-p_\perp^{(a)}/3]^3\approx [\pi~{\rm spectrum}]\sim \exp [-p_\perp^{(a)}/2]^2$, 
$p_\perp^{(a)}$ being a parton momentum; this is the essential point about the so called `coalescence' picture.

Attempts have been made to device models of fragmentation/coalescence (f/c)
and to calculate the $p_\perp$ dependence of certain experimental observables.
One of these observables, the so called `nuclear modification ratio', is:
\begin{center}
$
R_{AA}=1/N_{\rm coll}(b=0)\left[ (dN_H(b=0)/d^2p_\perp)|_{AA}/(dN_H/d^2p_\perp)|_{pp} \right]
$,
\end{center}
where $N_H$ is the number of hadrons counted, $b$ is the impact parameter of the
heavy-ion collision  ($b=0$ means maximum centrality), $AA$ labels  nucleus-nucleus collision ($pp$ for proton-proton) and $N_{\rm coll}$ is the number of nucleon-nucleon collisions occurred in $AA$.
Such a quantity can be measured experimentally and calculated in a f/c
model. The results are given in Fig.~1.

\begin{figure}
\centering
\includegraphics[height=4cm]{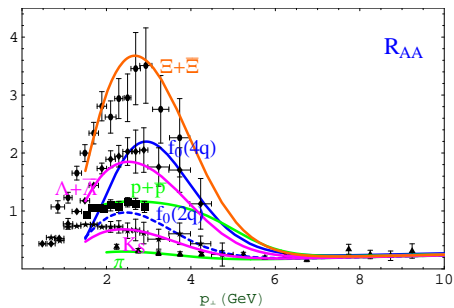}
%\picplace{5cm}{2cm} % Give the correct figure height and width in cm
%
\caption{The $R_{AA}$ value as a function of $p_\perp$ for various hadrons. The solid lines are the theoretical results obtained  in the f/c model.}
\label{fig:1}       % Give a unique label
\end{figure}
 
As shown, $R_{AA}$ has the ability to discriminate between mesons and baryons, as baryons tend to be higher in $R_{AA}$ than mesons. The curves are instead the result of a theoretical calculation in a f/c model.

Let us consider here the case of the $f_0(980)$ scalar meson which also evades any standard $q\bar q$ interpretation. Two possibilities are examined: the $f_0$ is (1) a $q\bar q$, (2) a diquark-antidiquark meson (a molecular picture in which the $f_0$ is a kind of $K\bar K$ molecule is as old as the discovery of the $f_0$ itself). 
The $R_{AA}(f_0)$ at RHIC has not yet been analyzed. We provide a couple of theoretical curves to eventually compare to data.
%Other similar observables can be studied with analogous techniques and results.
The $X$ will be produced at the LHC where an $R_{AA}(X)$ analysis might be performed.

\section{Conclusions}
It would be an error if collaborations like BELLE and BaBar gave up 
the investigation of a possible $\Delta M\neq 0$ between the $X$ produced in $B^0$ and $B^+$, or the search for charged $X$'s. Clarifying the nature of the $X$ and its `similes' gives an opportunity to learn some new fundamental aspects of quantum chromodynamics.
\subparagraph{Acknowledgements} I whish to thank L.~Maiani for fruitful and enjoyable collaboration and R.L.~Jaffe and R.~Faccini for many stimulating discussions. I conclude by thanking the organizers O.~Nicrosini, G.~Montagna and C.~Vercesi for their kind invitation and their valuable work.

%
% BibTeX users please use
% \bibliographystyle{}
% \bibliography{}
%
% Non-BibTeX users please follow the syntax
% the syntax of "referenc.tex" for your own citations
%%%%%%%%%%%%%%%%%%%%%%%% referenc.tex %%%%%%%%%%%%%%%%%%%%%%%%%%%%%%
% sample references
% "physics"
%
% Use this file as a template for your own input.
%
%%%%%%%%%%%%%%%%%%%%%%%% Springer-Verlag %%%%%%%%%%%%%%%%%%%%%%%%%%

%
% BibTeX users please use
% \bibliographystyle{}
% \bibliography{}

\begin{thebibliography}{99.}
% Use the following syntax and markup for your references
%
% Monographs
%\bibitem{monograph} H. Ibach, H. L\"uth: \textit{Solid-State
%Physics}, 2nd edn (Springer, Berlin Heidelberg New York 1996) pp 45--56

% Contributed Works
%\bibitem{contribution} D.M. MacKay: Visual stability and voluntary eye
%movements. In: \textit{Handbook of Sensory Physiology}, vol 3, ed by R.
%Jung, D.M. MacKay (Springer, Berlin Heidelberg New York 1973) pp
%307--331

% Journal
\bibitem{journal} {\it The X discovery}:
  S.~K.~Choi {\it et al.}  [Belle],
%   ``Observation of a new narrow charmonium state in exclusive 
%$B^{\pm} \to K^{\pm}  \pi^+\pi^- J/\psi$ decays,''
  Phys.\ Rev.\ Lett.\  {\bf 91}, 262001 (2003)
  [arXiv:hep-ex/0309032]; D.~Acosta {\it et al.}  [CDF II],
%   ``Observation of the narrow state $X(3872) \to J/\psi \pi^+ \pi^-$ in
% $\bar{p}p$  collisions at $\sqrt{s} = 1.96$ TeV,''
  Phys.\ Rev.\ Lett.\  {\bf 93}, 072001 (2004); 
  V.~M.~Abazov {\it et al.}  [D0],
%  ``Observation and properties of the $X(3872)$ decaying to $J/\psi \pi^+
% \pi^-$ in $p\bar{p}$ collisions at $\sqrt{s} = 1.96$ TeV,''
  Phys.\ Rev.\ Lett.\  {\bf 93}, 162002 (2004); 
  B.~Aubert {\it et al.}  [BABAR],
%   ``Study of the $B \to J/\psi K^- \pi^+ \pi^-$ decay and measurement 
% of the $B
%\to X(3872) K^-$ branching fraction,''
  Phys.\ Rev.\ D {\bf 71}, 071103 (2005).
\bibitem{journal} {\it Molecules}: 
  M.~B.~Voloshin and L.~B.~Okun,
  %``Hadron Molecules And Charmonium Atom,''
  JETP Lett.\  {\bf 23}, 333 (1976)
  [Pisma Zh.\ Eksp.\ Teor.\ Fiz.\  {\bf 23}, 369 (1976)];
  %%CITATION = JTPLA,23,333;%%
  N.~A.~Tornqvist,
  %``Possible large deuteron - like meson meson states bound by pions,''
  Phys.\ Rev.\ Lett.\  {\bf 67}, 556 (1991);
  %%CITATION = PRLTA,67,556;%%
  F.~E.~Close and P.~R.~Page,
  %``The D*0 anti-D0 threshold resonance,''
  Phys.\ Lett.\ B {\bf 578}, 119 (2004)
  [arXiv:hep-ph/0309253];
  %%CITATION = HEP-PH 0309253;%%
  E.~S.~Swanson,
  %``Short range structure in the X(3872),''
  Phys.\ Lett.\ B {\bf 588}, 189 (2004)
  [arXiv:hep-ph/0311229];
  %%CITATION = HEP-PH 0311229;%%
  E.~Braaten and M.~Kusunoki,
  %``Decays of the X(3872) into J/psi and light hadrons,''
  Phys.\ Rev.\ D {\bf 72}, 054022 (2005)
  [arXiv:hep-ph/0507163];
  %%CITATION = HEP-PH 0507163;%%
  M.~Suzuki,
  %``The X(3872) boson: Molecule or charmonium,''
  Phys.\ Rev.\ D {\bf 72}, 114013 (2005)
  [arXiv:hep-ph/0508258];
  %%CITATION = HEP-PH 0508258;%%
  E.~S.~Swanson,
  %``The new heavy mesons: A status report,''
  Phys.\ Rept.\  {\bf 429}, 243 (2006)
  [arXiv:hep-ph/0601110].
  %%CITATION = HEP-PH 0601110;%%
\bibitem{journal} {\it Diquarks $\&$ Tetraquarks}:
  R.~L.~Jaffe and F.~Wilczek,
  %``Diquarks and exotic spectroscopy,''
  Phys.\ Rev.\ Lett.\  {\bf 91}, 232003 (2003)
  [arXiv:hep-ph/0307341];
  %%CITATION = HEP-PH 0307341;%%
L.~Maiani, F.~Piccinini, A.~D.~Polosa and V.~Riquer,
  %``A new look at scalar mesons,''
  Phys.\ Rev.\ Lett.\  {\bf 93}, 212002 (2004)
  [arXiv:hep-ph/0407017];
  %%CITATION = HEP-PH 0407017;%%
Phys.\ Rev.\ D {\bf 70}, 054009 (2004)
  [arXiv:hep-ph/0407025];
Phys.\ Rev.\ D {\bf 71}, 014028 (2005)
  [arXiv:hep-ph/0412098];
Phys.\ Rev.\ D {\bf 72}, 031502 (2005)
  [arXiv:hep-ph/0507062];
AIP Conf.\ Proc.\  {\bf 814}, 508 (2006)
  [arXiv:hep-ph/0512082]; arXiv:hep-ph/0604018;
H.~Hogaasen, J.~M.~Richard and P.~Sorba,
  %``A chromomagnetic mechanism for the X(3872) resonance,''
  Phys.\ Rev.\ D {\bf 73}, 054013 (2006)
  [arXiv:hep-ph/0511039]; R.~Jaffe,
  %``Color non-singlet spectroscopy,''
  Phys.\ Rev.\ D {\bf 72}, 074508 (2005)
  [arXiv:hep-ph/0507149]; M.~Karliner and H.~J.~Lipkin,
%   ``Diquarks and antiquarks in exotics: A menage a trois and a menage a
% quatre,''
  Phys.\ Lett.\ B {\bf 638}, 221 (2006)
  [arXiv:hep-ph/0601193];
  R.~D.~Matheus, S.~Narison, M.~Nielsen and J.~M.~Richard,
  %``Can the X(3872) be a 1++ four-quark state?,''
  arXiv:hep-ph/0608297. 
  \bibitem{journal}
  {\it Lattice and Diquarks}: C.~Alexandrou, Ph.~de Forcrand and B.~Lucini,
  %``Evidence for diquarks in lattice QCD,''
  arXiv:hep-lat/0609004. 
  \bibitem{journal} {\it Hybrid}: F.~E.~Close and P.~R.~Page,
  %``The Production and decay of hybrid mesons by flux tube breaking,''
  Nucl.\ Phys.\ B {\bf 443}, 233 (1995)
  [arXiv:hep-ph/9411301]; E.~Kou and O.~Pene,
  %``Suppressed decay into open charm for the Y(4260) being an hybrid,''
  Phys.\ Lett.\ B {\bf 631}, 164 (2005)
  [arXiv:hep-ph/0507119]; F.~E.~Close and P.~R.~Page,
  %``Gluonic charmonium resonances at BaBar and Belle?,''
  Phys.\ Lett.\ B {\bf 628}, 215 (2005)
  [arXiv:hep-ph/0507199].
\bibitem{journal} {\it Counting Quarks}:
 L.~Maiani, A.~D.~Polosa, V.~Riquer and C.~A.~Salgado,
  %``Counting valence quarks at RHIC and LHC,''
  arXiv:hep-ph/0606217.
\end{thebibliography}

%%%%%%%%%%%%%%%%%%%%%%%%%%%%%%%%%%%%%%%%%%%%%%%%%%%%%%%%%%%%%%%%%%%%%%  }

%%%%%%%%%%%%%%%%%%%%%%%%%%%%%%%%%%%%%%%%%%%%%%%%%%%%%%%%%%%%%%%%%%%%%%

\printindex
\end{document}